# I Know What You Saw Last Minute - Encrypted HTTP Adaptive Video Streaming Title Classification

**Ran Dubin · Amit Dvir · Ofir Pele · Ofer Hadar**

**Abstract** Desktops and laptops can be maliciously exploited to violate privacy. There are two main types of attack scenarios: active and passive. In this paper, we consider the passive scenario where the adversary does not interact actively with the device, but he is able to eavesdrop on the network traffic of the device from the network side. Most of the Internet traffic is encrypted and thus passive attacks are challenging. Previous research has shown that information can be extracted from encrypted multimedia streams. This includes video title classification of non HTTP adaptive streams (non-HAS). This paper presents an algorithm for *encrypted HTTP adaptive video streaming title classification*. We show that an external attacker can identify the video title from video HTTP adaptive streams (HAS) sites such as YouTube. To the best of our knowledge, this is the first work that shows this. We provide a large data set of 10000 YouTube video streams of 100 popular video titles (each title downloaded 100 times) as examples for this task. The dataset was collected under real-world network conditions. We present several machine algorithms for the task and run a through set of experiments, which shows that our classification accuracy is more than 95%. We also show that our algorithms are able to classify video titles that are not in the training set as unknown and some of the algorithms are also able to eliminate false prediction of video titles and instead report unknown. Finally, we evaluate our algorithms robustness to delays and packet losses at test time and show that a solution that uses SVM is the most robust against these changes given enough training data. We provide the dataset and the crawler for future research.

Ran Dubin
Communication Systems Engineering
Ben-Gurion University of the Negev
Israel
Tel.: +972-8-6472592
Fax: +972-8-6472883
E-mail: dubinr@post.bgu.ac.il

Amit Dvir
Center for Cyber Technologies
Department of Computer Science
Ariel University
Israel
E-mail: amitdv@g.ariel.ac.il

Ofir Pele
Center for Cyber Technologies
Department of Computer Science
Department of Electrical and Electronics Engineering
Ariel University
Israel
E-mail: ofir.pele@g.ariel.ac.il

Ofer Hadar
Communication Systems Engineering
Ben-Gurion University of the Negev
Israel
E-mail: hadar@bgu.ac.il

## 1 Introduction

Every day, hundreds of millions of Internet users view videos online, whose numbers are clearly going to increase [1, 2]. By 2020, the share of video traffic is expected to increase to 82% of the total IP traffic, up from 70% in 2015. Google's streaming service, YouTube, now occupies a market share of over 17% of the total mobile network bandwidth in North America [2, 3].

Currently, most of the video streaming web sites including YouTube are using HTTP Adaptive Streaming (HAS). Dynamic Adaptive Streaming over HTTP (DASH) [4] is the *de facto* standard method for HAS. DASH is a Multi Bit Rate (MBR) streaming method that was designed to improve viewer Quality of Experience (QoE) [5]. In DASH, each video is divided into short segments, typically a few seconds long ($2-16$ seconds), and each segment is encoded several times, each time with a different quality representation. The user (player) Adaptation Logic (AL) algorithm is responsible for the automatic selection of the most suitable quality representation for each segment, based on the parameters such as client playout buffer and network conditions. As a result, the quality representation in DASH can change between segments.

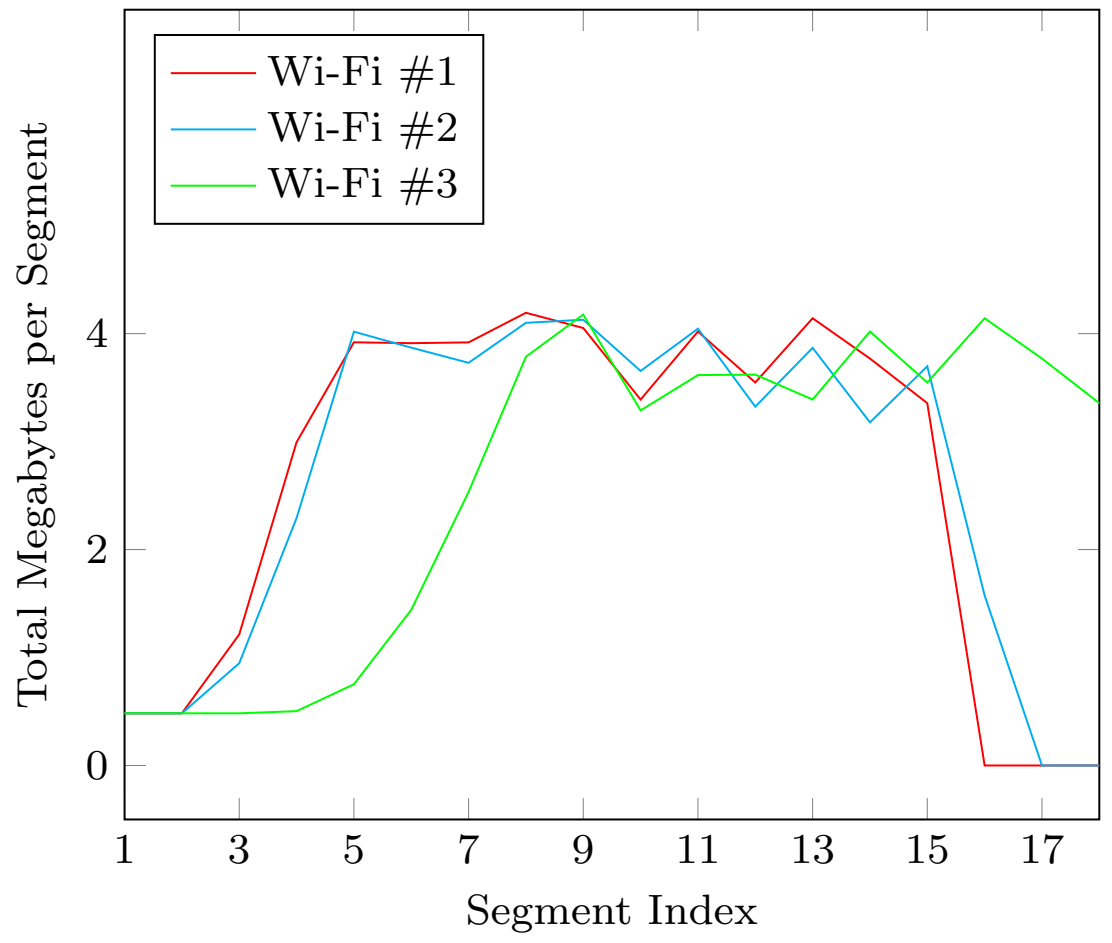

Fig. 1: Total megabytes per segment of three downloads over different Wi-Fi networks of the same video title, all with the same quality representation. Due to network conditions variability, there are differences between the networks.

In DASH, each quality representation is encoded in variable bit rates (VBRs). VBR varies the amount of output data per time segment and does not attempt to control the output bit rate of the encoder, so that the distortion will not vary significantly [6].

DASH often uses HTTP byte range mode. In this mode, the byte range of each segment request can be different. This depends on the client's network conditions and playout buffer levels. Fig. 1 shows an example for three downloads of the same video title over different Wi-Fi networks, all with the same quality representation. From the figure we can notice that due to networks conditions variability, there are differences between the networks.

YouTube analysis was conducted in many aspects such as YouTube server location [7, 8], comparison between YouTube and other video sharing services [7], PC *vs.* mobile user access patterns [9], QoE [10], traffic characterization and its DASH implementation [11] and network analysis [12–16].

YouTube has started to encrypt their video services [17]. As a result, traditional Deep Packet Inspection (DPI) methods for information retrieval in general and video title classification in particular are not viable.

Many recent works have suggested methods for encrypted traffic classification and several surveys have presented detailed descriptions of the state of the art methods [18–32]. Several works have examined different statistical features such as session duration [33–35], number of packets in a session [34,36,37], different variance calculations of the minimum, maximum and average values of inter-arrival packet time [34, 36], payload size information [36,38], bit rate [38,39], Round-Trip Time (RTT) [39], packet direction [40] and server sent bit rate [41]. All these features are not suitable for video streaming classification as the payload size in video streaming is often maximum size, delays in the network are varied, and re-transmissions cause false packet counts.

Recent works showed that video title classification of encrypted video streams is possible [28–30]. These works use features such as packet size and the application layer information. Saponas et al. [28] uncovered security issues with consumer electronic gadgets that enable information retrieval such as video title classification. Liu et al. [29] presented a method for video title classification of RTP/UDP internet traffic. In [30] Liu et al. presented an improved algorithm which is more efficient and demonstrated excellent results on a bigger data set with real network conditions. They used the wavelet transform for constructing unique and robust video signatures with different compactnesses.

Since these works [28–30] were conducted, there have been several changes in video traffic over the internet: (i) Adaptive byte range selection over HTTP; (ii) MBR adaptive streaming; (iii) HTTP version 2 [42]. This paper's main contributions are:

- This is the first work that shows that a passive attacker, sniffing ISP or Wi-Fi open-system network traffic, can identify video titles of encrypted YouTube video streams over DASH. Inspired by other works presented above, we exploit traffic patterns and Variable Bit-Rate (VBR) encoding. We present new methods that are applicable also to current standards of video streaming.
- We run through a set of experiments, which shows that our classification accuracy is more than 95%.
- We show that our algorithms are able to classify video titles that are not in the training as unknown and some of the algorithms are also able to eliminate false prediction of video titles and instead report unknown.
- We evaluate our algorithms robustness to delays and packet losses at testing time and show that a solution that uses SVM is the most robust against these changes given enough training data.
- We provide a comprehensive dataset that contains 10000 labeled YouTube streams of 100 video titles (that is, 100 streams per video title). The streams were downloaded from the Internet under real-world network conditions. The dataset [43] and crawler [44] are available for download.

The remainder of this paper is organized as follows. In Section 2 we present our framework and our suggested algorithms. In Section 3 we evaluate the performance of all algorithms also under severe network conditions on testing times. Finally, we conclude in Section 4.

## 2 Video Title Classification

The proposed solution architecture has three modules. The first module (Section 2.1) removes non-YouTube packets and audio packets. The next module combines several YouTube packets into a *peak*. A peak is defined as a section of traffic where there is silence before and after. Features are extracted from these peaks (Section 2.2) and passed into a classification algorithm (Section 2.3). It is noteworthy that the input to all of our classification algorithms is only encrypted HTTP adaptive video streaming traffic.

### 2.1 Preprocessing

First, we divide the traffic into flows based on a five-tuple representation: {protocol (TCP/UDP), src IP, dst IP, src port, dst port}. Then, we decide for each flow whether it is a YouTube flow. This is done based on the Service Name Indication (SNI) field in the *Client Hello* message. If the "*googlevideos.com*" string is found in the SNI, the flow is passed to the next module. Note that the YouTube flows identification can also be done using machine learning techniques [45, 46].

Second, we optionally remove audio packets. In all our training data, bursts that were smaller than 400kB, while video traffic bursts were much larger. The audio data and the video data can be found in the same 5-tuple flow and in some cases we cannot distinguish between them.

Finally, we remove TCP re-transmissions using a TCP stack [47] as re-transmissions are caused mostly by network conditions.

### 2.2 Feature Extraction

The feature extraction is done on the preprocessed traffic, where non-YouTube flows, audio packets, and TCP re-transmissions have been removed. To better understand encrypted YouTube streaming traffic properties, we examined YouTube traffic under different browsers. Fig. 2 depicts traffic download patterns of auto quality representation using different browsers. In the figure we can see that all flows contain peaks. Rao et al. [48] and Ameigeiraset al. [49] showed the same characteristics (coined in [48, 49] as "On/Off"). Therefore we decided to encode every peak of the stream to a feature. This feature is the Bit-Per-Peak (BPP); that is, total number of bits in a peak.

### 2.3 Machine Learning

After the preprocessing and feature extraction, each video stream (number $j$ of video title $i$) is represented by $\mathcal{S}_{ij}$, a set of Bit-Per-Peak (BPP) features (no duplicates). Note that each BPP-set may have different cardinality.

We adapt four machine learning algorithms. The first is the nearest neighbor algorithm [50]. In this algorithm, a testing stream title is determined as the nearest neighbor stream title in the training data. The second and third are the nearest neighbor to class algorithm [51] and the nearest neighbor to class unique algorithm.

These algorithms compute the similarity score between a test sample and *all* training samples of a class (video title). In the unique version only features that appear in all streams of a video title are used. The fourth algorithm generalizes the second, by using similarities as features [52] in an SVM algorithm [53]. Table 1 summarizes symbols used in this paper. Following are detailed explanations of our adaptations of the machine learning algorithms.

#### *2.3.1 Nearest Neighbor (NN) Algorithm*

The nearest neighbor similarity score between two BPP-sets, $\mathcal{S}$ and $\mathcal{S}'$, is the cardinality of the intersection set:

$$\mathrm{sim}(\mathcal{S}, \mathcal{S}') = |\mathcal{S} \cap \mathcal{S}'| \tag{1}$$

At test time, each video stream BPP-set, $\mathcal{S}_{\mathrm{test}}$, is classified as the video title $i$, that has the maximum similarity score to one of the title training stream BPP-sets or as unknown if all similarities are zero:

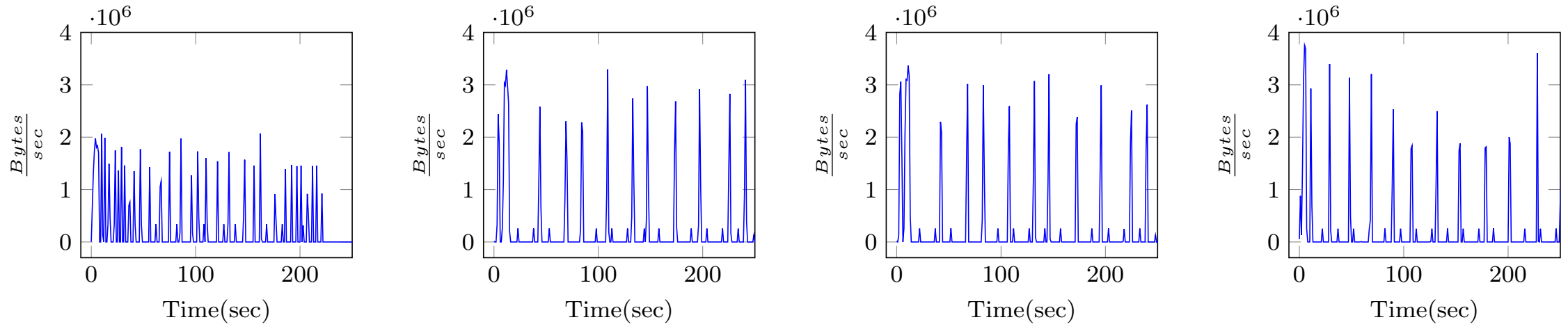

(a) Firefox (b) Safari (c) Explorer (d) Chrome

Fig. 2: Traffic flows of auto mode downloads of the same movie from different browsers. All flows have the same characteristics: *peaks* (of packets) with silences before and after. Note that the differences between the flows may be caused by: auto mode, network conditions, video container, video encoder, etc.

| | |
|---|---|
| BPP | Bit-Per-Peak |
| $i$ | Video title number |
| $j$ | Stream number |
| $n$ | Number of video titles in the training dataset |
| $m_i$ | Number of stream samples per title $i$ in the training dataset |
| $\mathcal{S}, \mathcal{S}'$ | BPP-sets |
| $\mathcal{S}_{ij}$ | A BPP-set of stream number $j$ of title $i$ |
| $\mathcal{S}_{\text{test}}$ | A test BPP-set |
| $\mathcal{T}_i$ | A BPP-set which is a union of all training streams of video $i$ (Eq. 4) |
| $\mathcal{U}_i$ | A BPP-set which is a union of all training streams of video $i$ minus BPPs of other video titles (Eq. 7) |

Table 1: List of Abbreviations

$$\forall \ 1 \le i \le n, \quad s_i = \max_{j=1}^{m_i} \text{sim}(\mathcal{S}_{\text{test}}, \mathcal{S}_{ij}) \tag{2}$$

$$y(\mathcal{S}_{\text{test}}) = \begin{cases} \operatorname*{argmax}_{i=1}^{n} s_i & \text{if } \left(\max_{i=1}^{n} s_i\right) \neq 0 \\ \text{unknown} & \text{otherwise} \end{cases} \tag{3}$$

### *2.3.2 Nearest Neighbor to Class (NNC) Algorithm*

In the nearest neighbor to the class algorithm, each video title $i$ in the training is represented by a single BPP-set, $\mathcal{T}_i$, which is a union of all its $m_i$ video stream BPP-sets (no duplicates):

$$\mathcal{T}_i = \cup_{j=1}^{m_i} \mathcal{S}_{ij} \tag{4}$$

As in the nearest neighbor algorithm, the similarity score is the cardinality of the intersection set. In this case, the similarity is between a BPP-set of a single stream and the BPP-set of all streams of a title:

$$\text{sim}(\mathcal{S}, \mathcal{T}_i) = |\mathcal{S} \cap \mathcal{T}_i| \tag{5}$$

At test time, each video stream set, $\mathcal{S}_{\text{test}}$, is classified as the video title $i$, that has the maximum similarity score to one of the $n$ video title BPP-sets or as unknown if all similarities are zero:

$$y(\mathcal{S}_{\text{test}}) = \begin{cases} \operatorname*{argmax}_{i=1}^{n} \text{sim}(\mathcal{S}_{\text{test}}, \mathcal{T}_i) & \text{if } \left(\max_{i=1}^{n} \text{sim}(\mathcal{S}_{\text{test}}, \mathcal{T}_i)\right) \neq 0 \\ \text{unknown} & \text{otherwise} \end{cases} \tag{6}$$

### *2.3.3 Nearest Neighbor to Class Unique (NNCU) Algorithm*

As in the nearest neighbor to class algorithm, in the nearest neighbor to class unique algorithm, each video title $i$ in the training is represented by a single BPP-set. In the nearest neighbor to class unique algorithm, the set is a union of all its $m_i$ video stream BPP-sets (no duplicates) *minus* BPP values that appear in sets of other video titles:

$$\mathcal{U}_i = \mathcal{T}_i \setminus \{\cup_{i'=1, i' \neq i}^{n} \mathcal{T}_{i'}\} \tag{7}$$

As in the nearest neighbor to class algorithm, the similarity score is the cardinality of the intersection set:

$$\text{sim}(\mathcal{S}, \mathcal{U}_i) = |\mathcal{S} \cap \mathcal{U}_i| \tag{8}$$

At test time, each video stream set, $\mathcal{S}_{\text{test}}$, is classified as the video title $i$, that has the maximum similarity score to one of the $n$ video title BPP-sets or as unknown if all similarities are zero:

$$y(\mathcal{S}_{\text{test}}) = \begin{cases} \operatorname*{argmax}_{i=1}^{n} \text{sim}(\mathcal{S}_{\text{test}}, \mathcal{U}_i) & \text{if } \left(\max_{i=1}^{n} \text{sim}(\mathcal{S}_{\text{test}}, \mathcal{U}_i)\right) \neq 0 \\ \text{unknown} & \text{otherwise} \end{cases}$$

(9)

#### 2.3.4 Similarities as Features Support Vector Machine (SFSVM) Algorithm

In this algorithm, each video stream is represented by a feature vector which is the video stream similarity to all $n$ video title sets (thus it is an $n$-dimensional vector). Where the similarity is the same as in the nearest neighbor to class algorithm, Eq. 5:

$$\vec{x}(\mathcal{S}) = [\text{sim}(\mathcal{S}, \mathcal{T}_1), \ldots, \text{sim}(\mathcal{S}, \mathcal{T}_n)] \tag{10}$$

Thus, the training set is an $(\sum_{i=1}^{n} m_i) \times n$ matrix of all training stream feature vectors:

$$\begin{bmatrix} \text{sim}(\mathcal{S}_{11}, \mathcal{T}_1) & \ldots & \text{sim}(\mathcal{S}_{11}, \mathcal{T}_n) \\ \vdots & \ddots & \vdots \\ \text{sim}(\mathcal{S}_{1m_1}, \mathcal{T}_1) & \ldots & \text{sim}(\mathcal{S}_{1m_1}, \mathcal{T}_n) \\ \vdots & \ddots & \vdots \\ \text{sim}(\mathcal{S}_{n1}, \mathcal{T}_1) & \ldots & \text{sim}(\mathcal{S}_{n1}, \mathcal{T}_n) \\ \vdots & \ddots & \vdots \\ \text{sim}(\mathcal{S}_{nm_n}, \mathcal{T}_1) & \ldots & \text{sim}(\mathcal{S}_{nm_n}, \mathcal{T}_n) \end{bmatrix} \tag{11}$$

We learn one vs. all support vector machines [53,54]. That is, we learn a classifier for each video title $i$ that classifies whether it is this title or any of the other titles. The classifiers are $n$-dimensional weight vectors and at test time, each video stream set, $\mathcal{S}_{\text{test}}$, is classified as the video title $i$, which maximizes the dot product between the class weight vector and the features vector or as unknown if all similarities are zero:

$$\vec{x}(\mathcal{S}_{\text{test}}) = [\text{sim}(\mathcal{S}_{\text{test}}, \mathcal{T}_1), \ldots, \text{sim}(\mathcal{S}_{\text{test}}, \mathcal{T}_n)] \tag{12}$$

$$y(\mathcal{S}_{\text{test}}) = \begin{cases} \underset{i=1}{\overset{n}{\text{argmax}}} \, (\vec{w_i} \cdot \vec{x}(\mathcal{S}_{\text{test}})) & \text{if } \left(\max_{i=1}^{n} \text{sim}(\mathcal{S}_{\text{test}}, \mathcal{T}_i)\right) \neq 0 \\ \text{unknown} & \text{otherwise} \end{cases} \tag{13}$$

It is noteworthy that if we learn the following weight vectors:

$$\forall \;\; 1 \leq i \leq n, \;\; \vec{w_i} = [0, \ldots, \overset{i}{\smile}{1}, \ldots, 0] \tag{14}$$

This exactly models the nearest neighbor to the class algorithm. Thus, this algorithm generalizes the nearest neighbor to the class algorithm.

A summary of the algorithms' training and testing samples is in Table 2.

## 3 Performance Evaluation

In this section, we evaluate the proposed encrypted HTTP adaptive video streaming title classification algorithms. First, we describe the dataset in 3.1. Then we report experimental results in Section 3.2.

### 3.1 Dataset

We collected a training set of encrypted video streams. The dataset contains 10000 labeled YouTube streams of 100 video titles (that is, 100 stream downloads per video title). The streams were downloaded from Youtube via the Internet (thus, each downloaded stream had different network conditions). The video titles used in this study are popular YouTube videos from different categories such as news, sports, nature, video action trailers, and GoPro videos. The dataset and crawler are available for download at [43].

In this study we decided to use the Chrome browser since it is the most popular browser in the market and its popularity is growing [55]. We used the default auto mode of the YouTube player (the player decides which quality representation to download based on estimation of the client network conditions).

We used the Selenium web automation tool [56] with ChromeDriver [57] for the crawler, so it will simulate a user video download. We used Adblock Plus [58] to eliminate advertisements.

### 3.2 Experimental Results

We recall that our classifiers have two type of predictions: a video title $1 \leq i \leq n$ and unknown. Unknown means that the classifier predicts that the given stream video title is not in the training set. We use the following evaluation metrics:

Accuracy Number of times that the classifier predicted video title $i$ and it was true, divided by the total number of predictions.

False-Prediction-Error Number of times that the classifier predicted video title $i$ and it was false, divided by the total number of predictions.

Unknown-True-Prediction Number of times that the classifier predicted video title unknown and it is indeed not a video title from the training set, divided by the total number of predictions.

Unknown-Prediction-Error Number of times that the classifier predicted video title unknown while it was a video title from the training set, divided by the total number of predictions.

| Algorithm | Training Sample | Testing Sample |
|---|---|---|
| *Nearest Neighbor* | $\mathcal{S}_{ij}$ | $\mathcal{S}_{\text{test}}$ |
| *Nearest Neighbor to Class* | $\mathcal{T}_i$ | $\mathcal{S}_{\text{test}}$ |
| *Nearest Neighbor to Class Unique* | $\mathcal{U}_i$ | $\mathcal{S}_{\text{test}}$ |
| *Similarities as Features Support Vector Machine* | $\vec{x}(\mathcal{S}_{ij}) = [\text{sim}(\mathcal{S}_{ij}, \mathcal{T}_1), \dots, \text{sim}(\mathcal{S}_{ij}, \mathcal{T}_n)]$ | $\vec{x}(\mathcal{S}_{\text{test}}) = [\text{sim}(\mathcal{S}_{\text{test}}, \mathcal{T}_1), \dots, \text{sim}(\mathcal{S}_{\text{test}}, \mathcal{T}_n)]$ |

Table 2: Algorithms' training and testing samples

We first report results using variable training dataset sizes. For all following experiments, we used 1000 streams (10 streams per video title) as the testing set. For training, we used the other 9000 streams (90 streams per video title), 6000 streams (60 streams per video title), 3000 streams (30 streams per video title) and 500 streams (5 streams per video title). All the test video streams were different from the ones that were used in the training phase, because of network conditions while streaming video from Youtube.

In these experiments, the testing set did not contain streams of video titles that were not in the training data. So, the Unknown-True-Prediction was 0%. We compared all algorithms with our features. We also experimented with and without the removal of audio features. The results of these experiments are depicted in Fig. 3.

There are several observations. First, all algorithms were able to accurately identify the video title of an encrypted HTTP adaptive stream (HAS). Accuracy was higher than 90% using 60 or more streams per video title. Even using only 5 streams per video title, NN+A, NNC+A and SFSVM+A accuracy was larger than 90%. Using also BPPs of audio peaks the accuracy was higher than 95% using 90 streams per video title, but the False-Prediction-Error which is a more severe error than the Unknown-Prediction-Error was also higher. The NNCU algorithm accuracy was lower in comparison to the other algorithms and moreover without the audio features. But, the accuracy is still high and the False-Prediction-Errors were almost eliminated.

We also experimented with 30 video titles that were not in the training set. The True-Unknown-Prediction rate (predicting unknown video title when it is not in the training dataset) was 100% for all algorithms. That is, all 30 video titles that were not in the training data were classified correctly as unknown.

As network conditions vary, we also tested our algorithms with additional LAN network delay and packet loss on test time. That is, only the testing data is changed and not training data. The additional delay and drop affects the client player and causes it to select different (lower) representations. The delays and packet loss were added using the clumsy application [59].

The results of additional LAN network delay are depicted in Fig. 4. We can see that using the largest training dataset the SFSVM+A method outperformed all other methods and achieved accuracy of more than 80% even under severe network delays of 600msec. We conjecture that the learning phase and the usage of distance to class made this algorithm robust to changes while downloading the video. We can see that NNC+A accuracy was also high. Training with less streams, the algorithms NNC+A and SFSVM+A are comparable and both outperformed all other algorithms. Like previous results the NNCU algorithm and not using audio features eliminated False-Prediction-Error. Out of all algorithms with low False-Prediciton-Error, NNCU+A accuracy was the best.

The results of additional packet loss are depicted in Fig. 5. We can see that using the largest training dataset the SFSVM+A method slightly outperformed all other methods and achieved accuracy of more than 70% even under severe packet loss of 6%. NNC+A, NNCU+A, NN+A all also had good performance. Training with less streams, methods accuracy was generally reduced. Similar to previous results the NNCU algorithm and not using audio features almost eliminated False-Prediction-Error. Again, out of all algorithms with low False-Prediciton-Error, NNCU+A accuracy was the best.

## 4 Conclusions

This paper showed that the video title of encrypted HTTP adaptive streams such as YouTube can be identified with high accuracy, even under severe network conditions. To the best of our knowledge this is the first work to show this. We presented several algorithms for this task and compared them on a large real-world traffic dataset. Overall, having enough training data the SFSVM+A algorithm achieved the best accuracy even under severe network conditions. If False-Prediciton-Error (predicting the wrong title) is a severe error we recommend to use the NNCU+A algorithm which instead reports unknown on almost all of its errors and still has relatively high accuracy. The dataset and the crawler are provided for future research at [43].

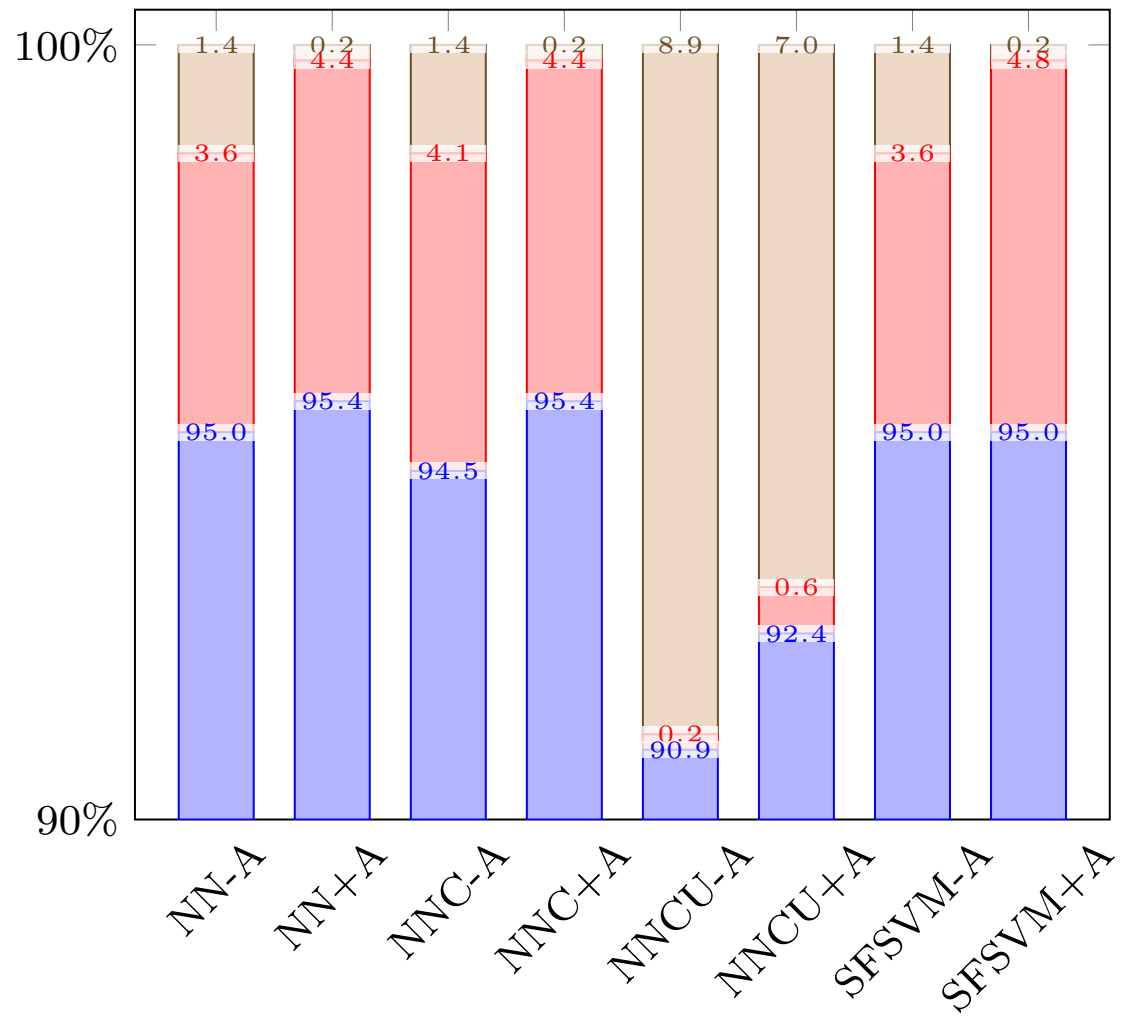

(a) 90 streams per video title in training (9000 streams total)

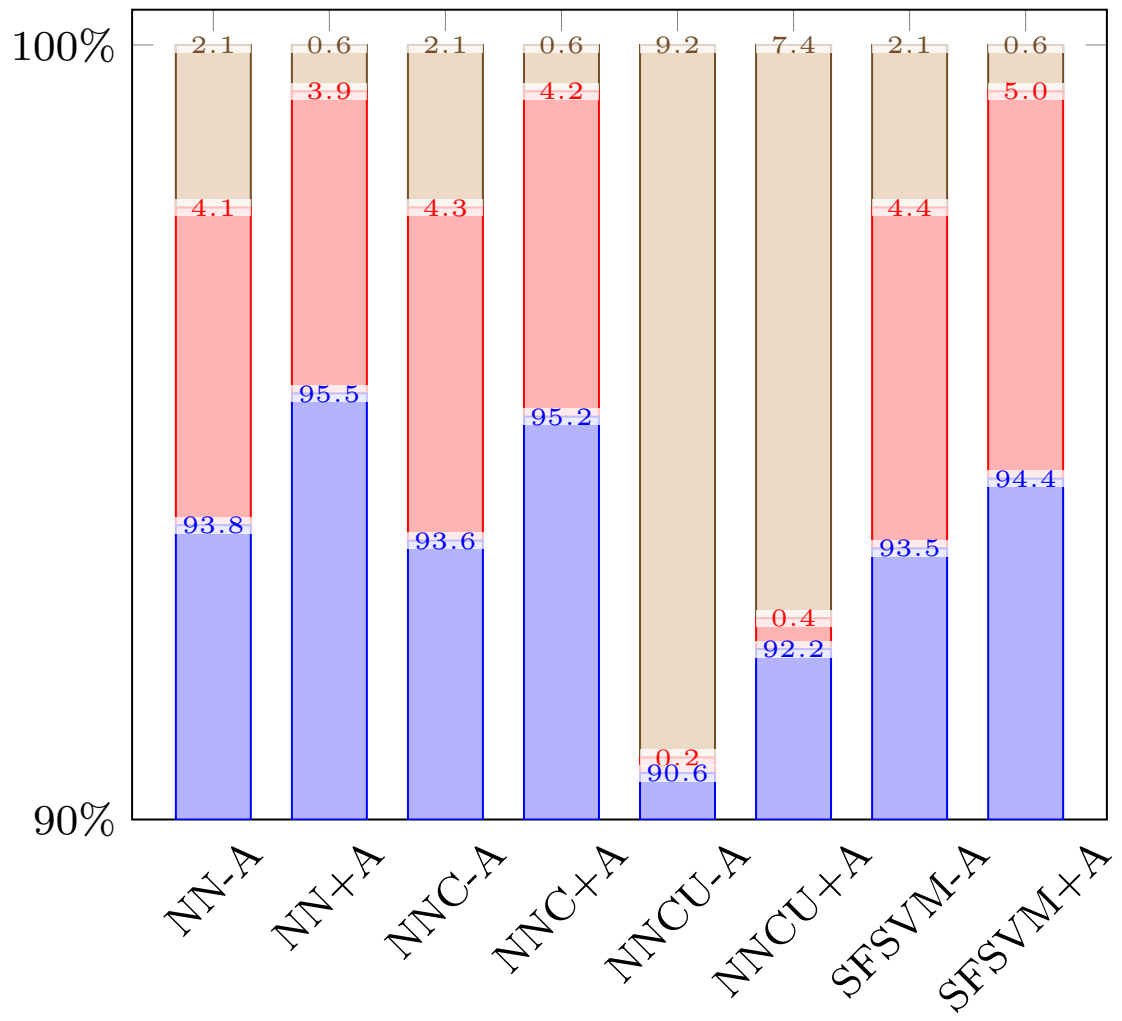

(b) 60 streams per video title in training (6000 streams total)

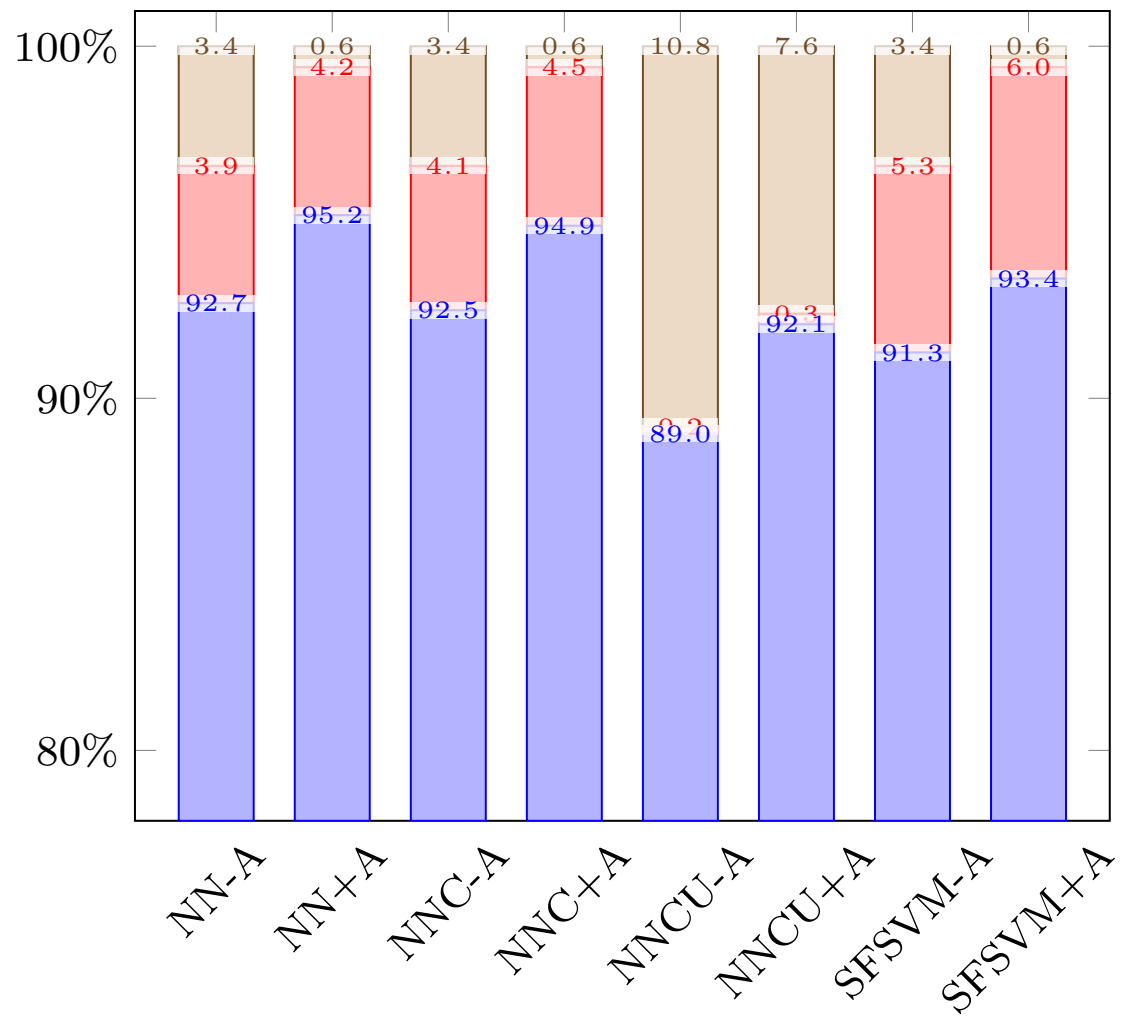

(c) 30 streams per video title in training (3000 streams total)

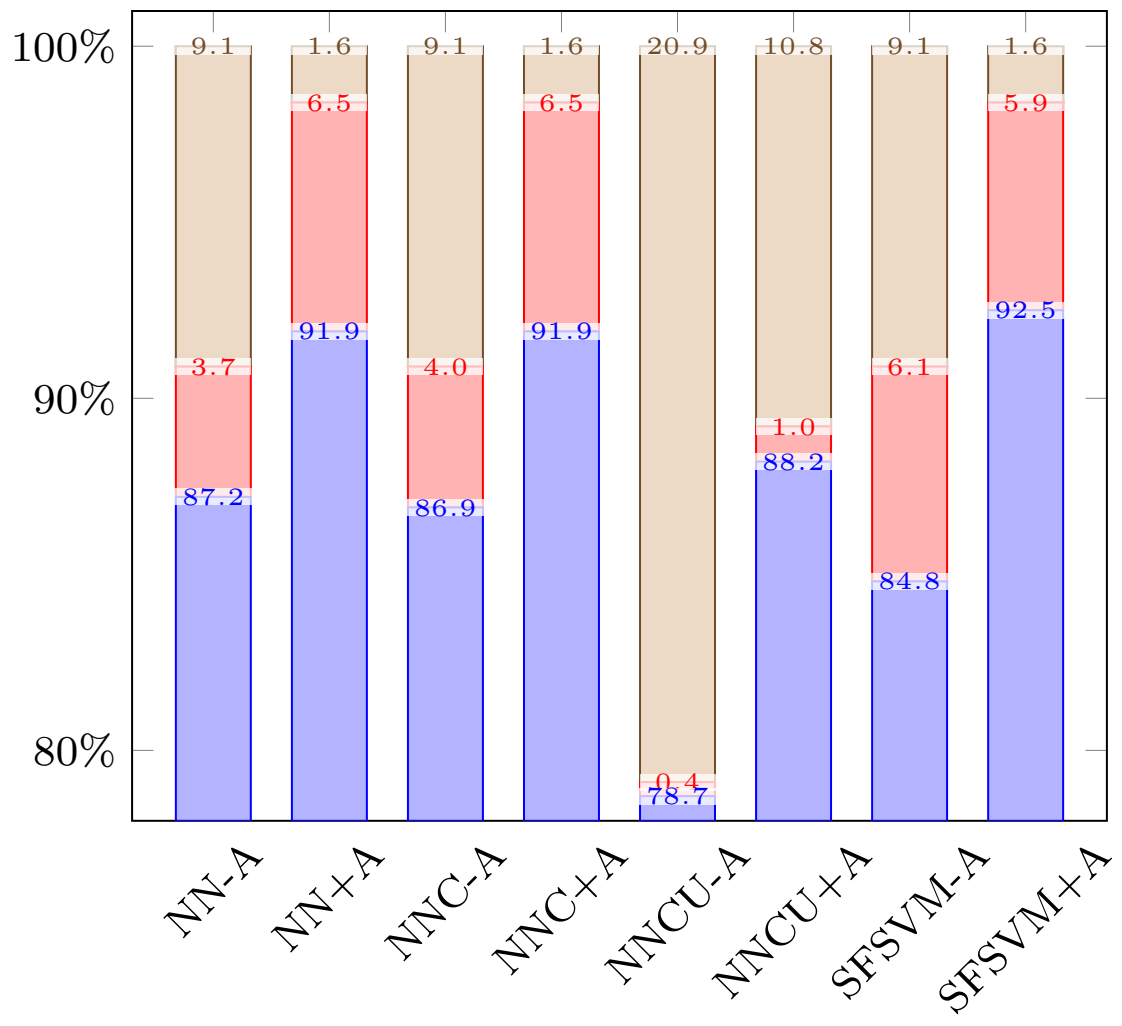

(d) 5 streams per video title in training (500 streams total)

- Unknown-Prediction-Error: predict video title unknown while it is video title $i$ from the training data
- False-Prediction-Error: predict video title $i$ while it is video title $i' \neq i$ from the training data
- Accuracy: predict the right video title

| | |
|---|---|
| NN: Nearest Neighbor | NNC: Nearest Neighbor to Class |
| NNCU: Nearest Neighbor to Class Unique | SFSVM: Similarities as Features Support Vector Machine |
| -A: without audio features | +A: with audio features |

Fig. 3: Accuracy, False-Prediction-Error and False-Unknown-Error (predicting unknown video title when it is in the training dataset) results for different training data set sizes and different learning algorithms. We can see that all algorithms were able to identify the video title of an encrypted HTTP adaptive stream (HAS) with very high accuracy. Using 60 or more streams per video title in the training data set, all algorithms achieved accuracy higher than 90% (as there are 100 classes a chance classifier accuracy is only 1% for this task). Even using only 5 streams per video title, NN+A, NNC+A and SFSVM+A achieved more than 90% accuracy. Adding audio features increased accuracy. However, usually it also increased the False-Prediction-Error which is a more severe error than the Unknown-Prediction-Error. The NNCU algorithm, especially without the audio features, achieved lower accuracy as compared to the other algorithms. However, the accuracy was still high (89% or higher for 30 streams per video title or more, and 78.7% for 5 streams per video title) and the False-Prediction-Error was almost eliminated.

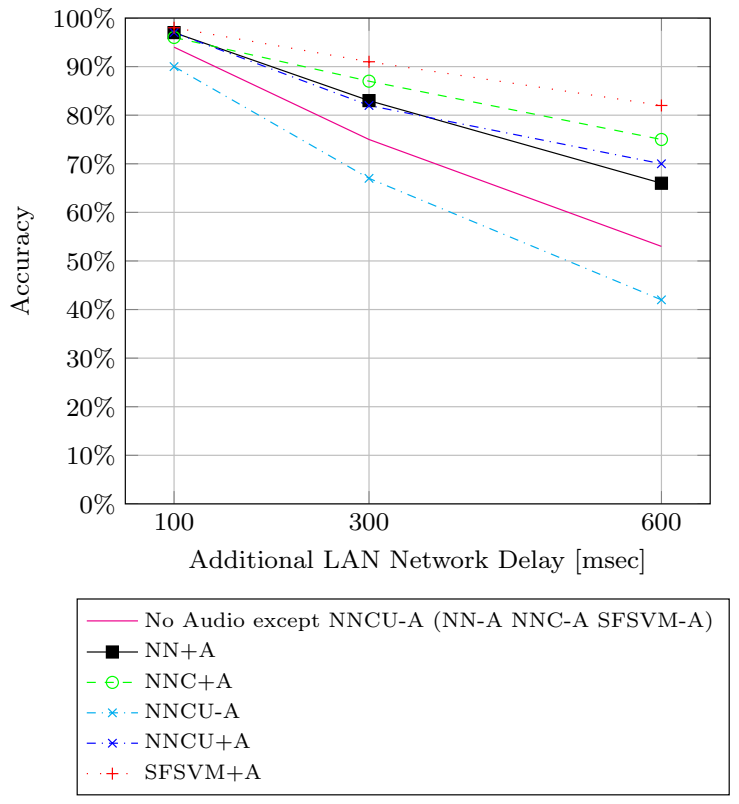

(a) Accuracy, 90 streams per video title in training (9000 streams total)

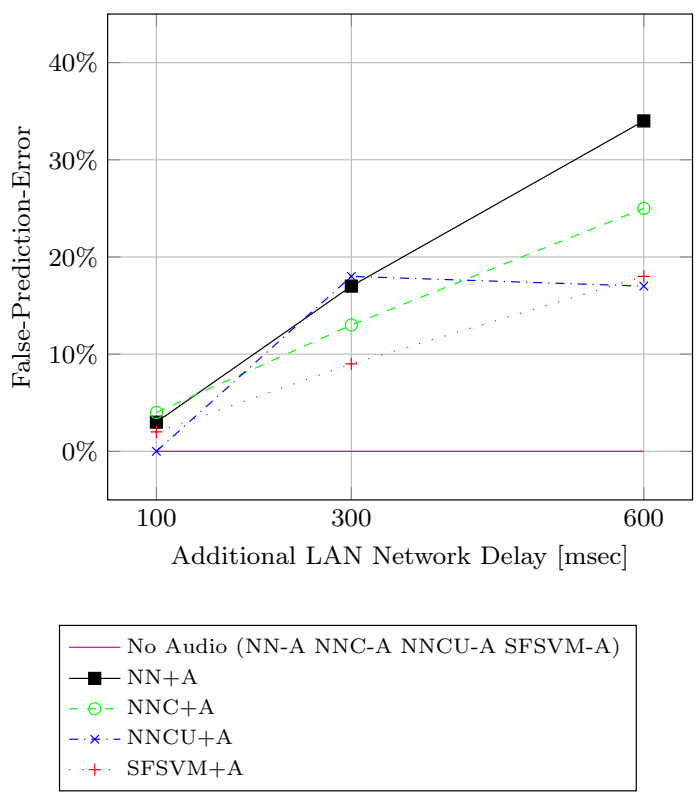

(b) False-Prediction-Error, 90 streams per video title in training (9000 streams total)

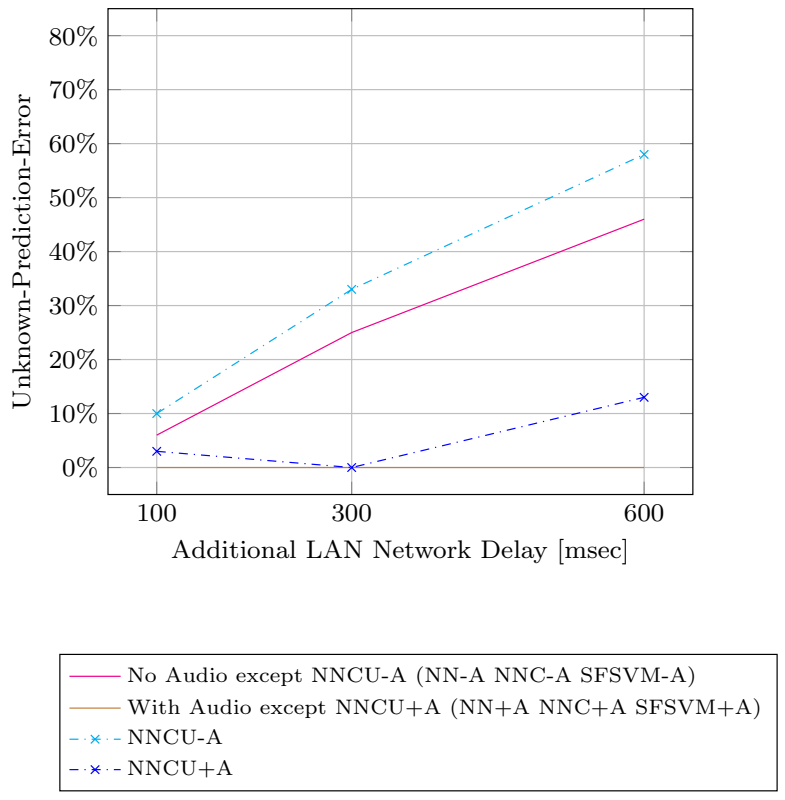

(c) Unknown-Prediction-Error, 90 streams per video title in training (9000 streams total)

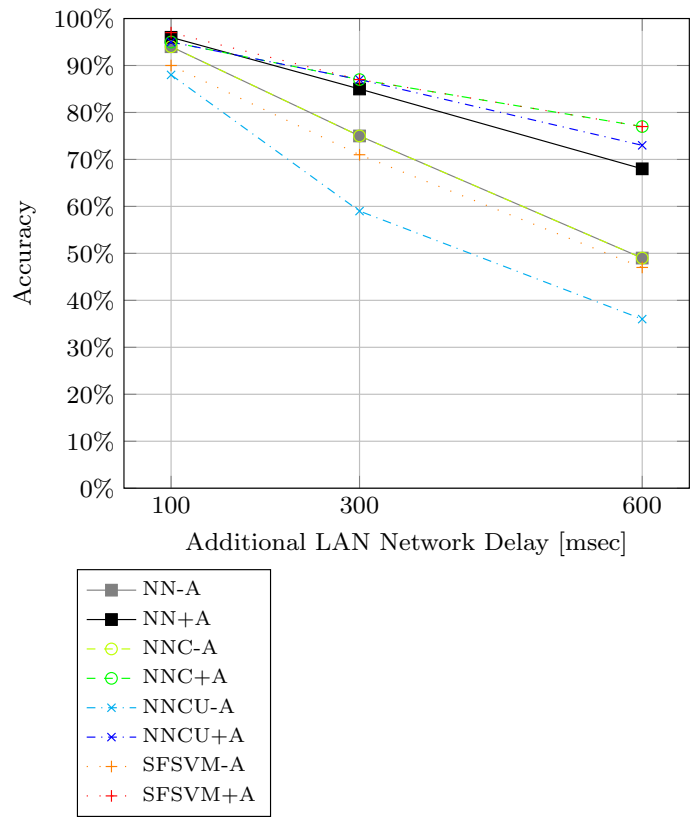

(d) Accuracy, 60 streams per video title in training (6000 streams total)

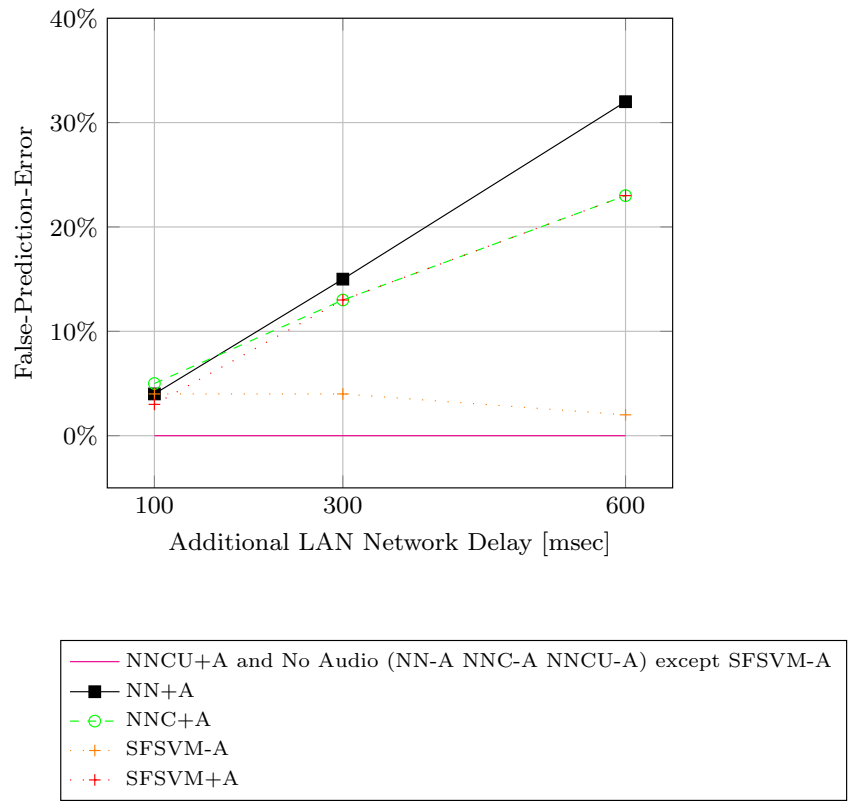

(e) False-Prediction-Error, 60 streams per video title in training (6000 streams total)

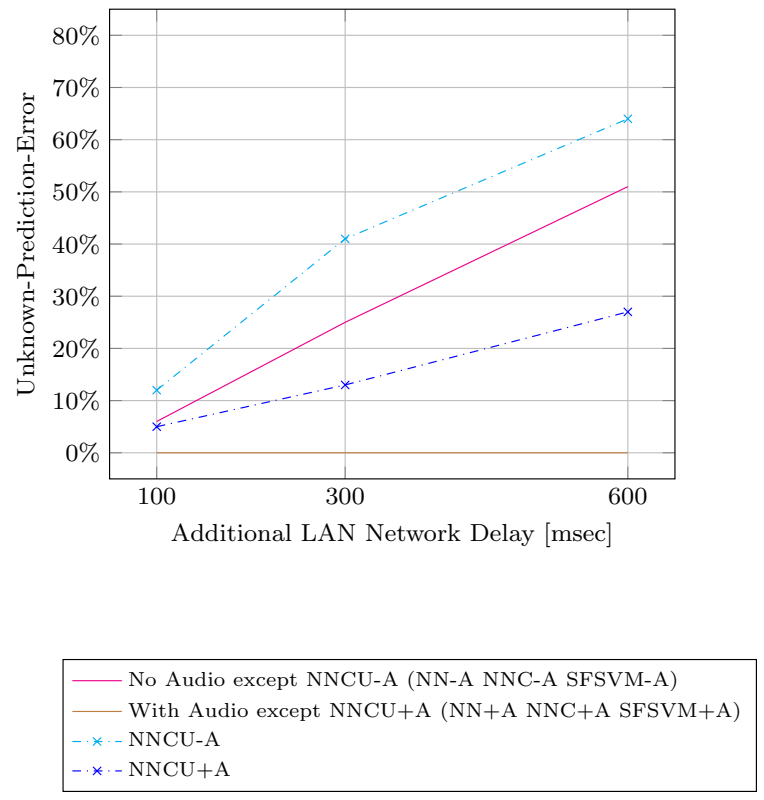

(f) Unknown-Prediction-Error, 60 streams per video title in training (6000 streams total)

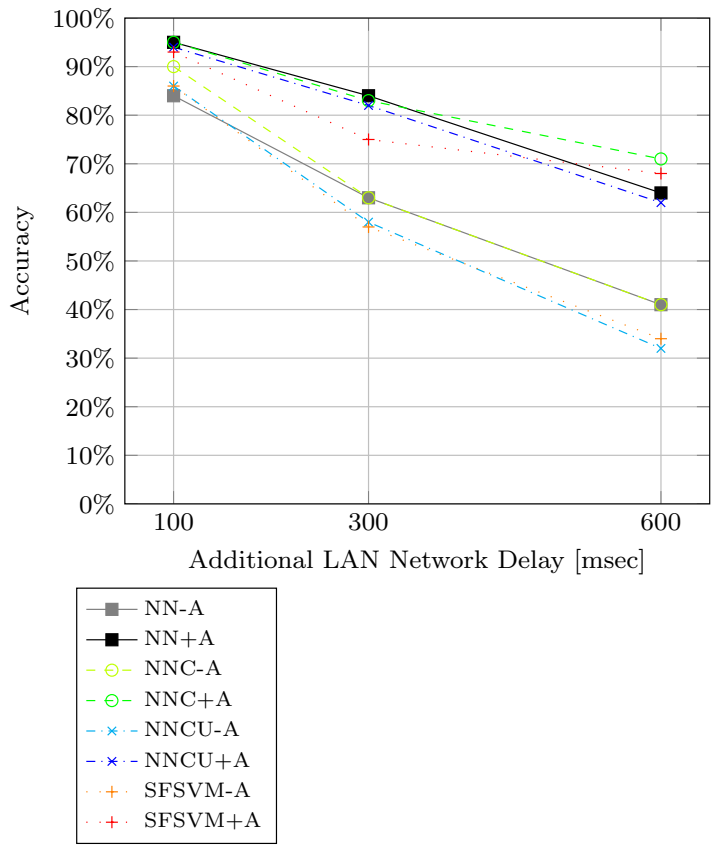

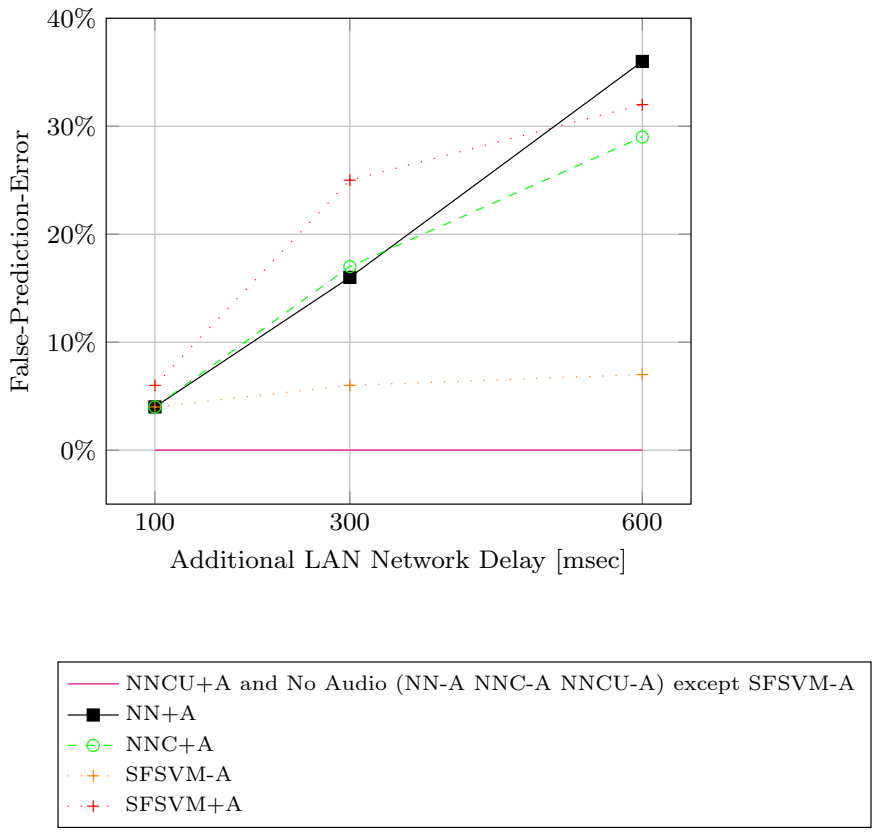

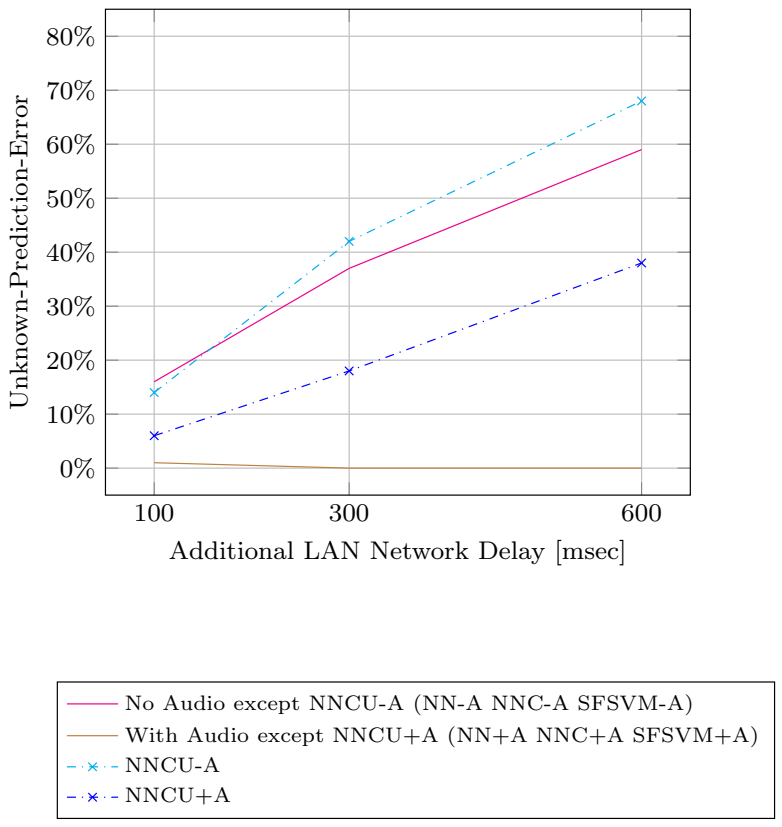

(g) Accuracy, 30 streams per video title in training (3000 streams total)

(h) False-Prediction-Error, 30 streams per video title in training (3000 streams total)

(i) Unknown-Prediction-Error, 30 streams per video title in training (3000 streams total)

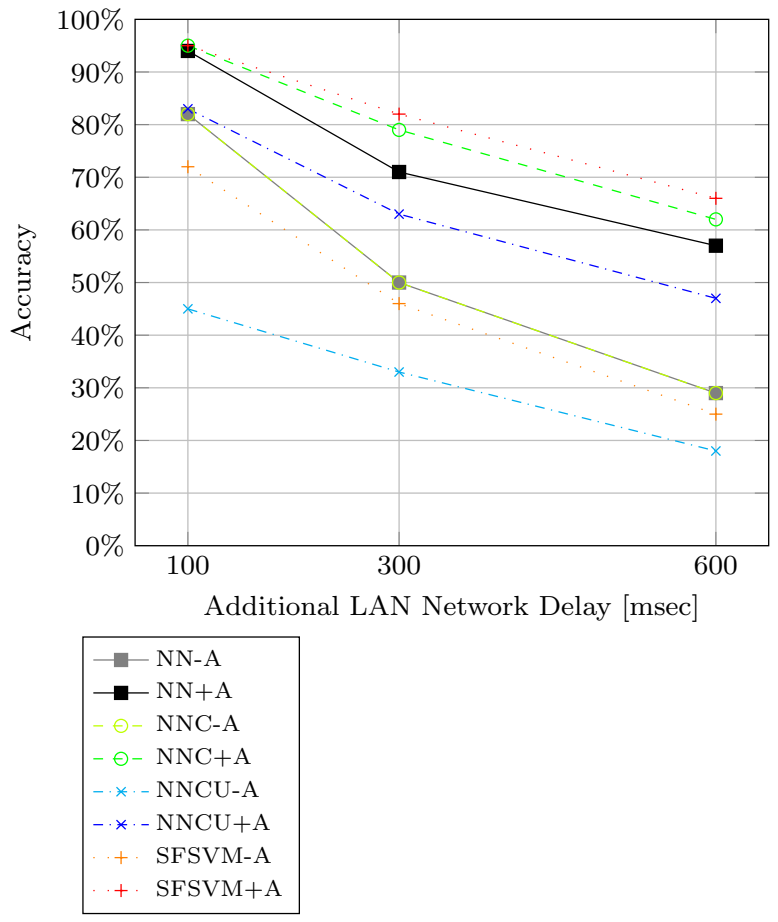

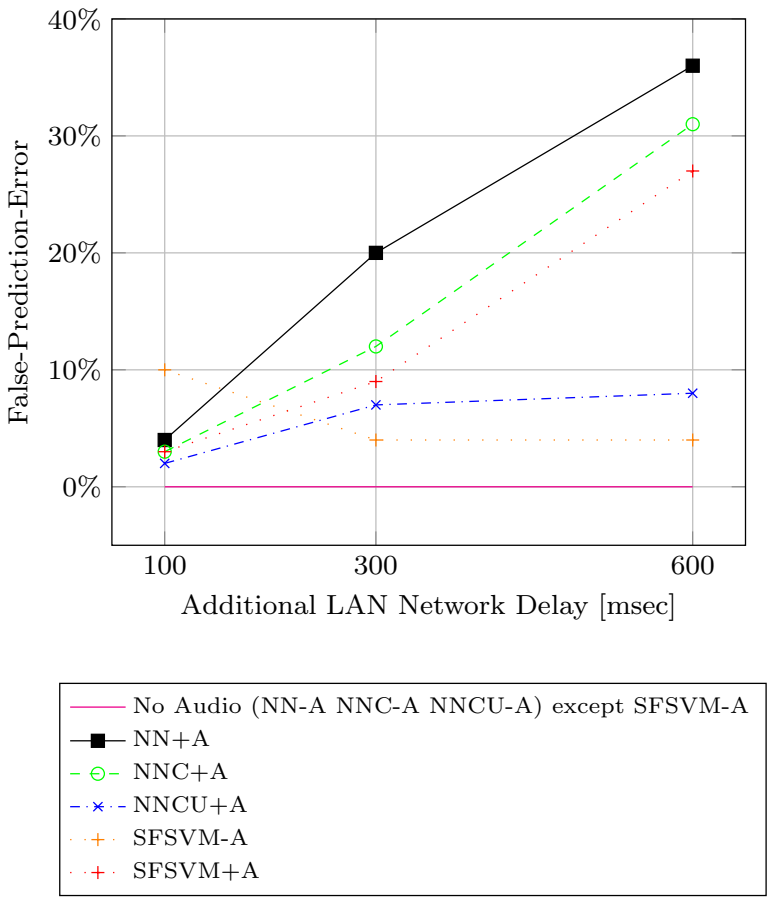

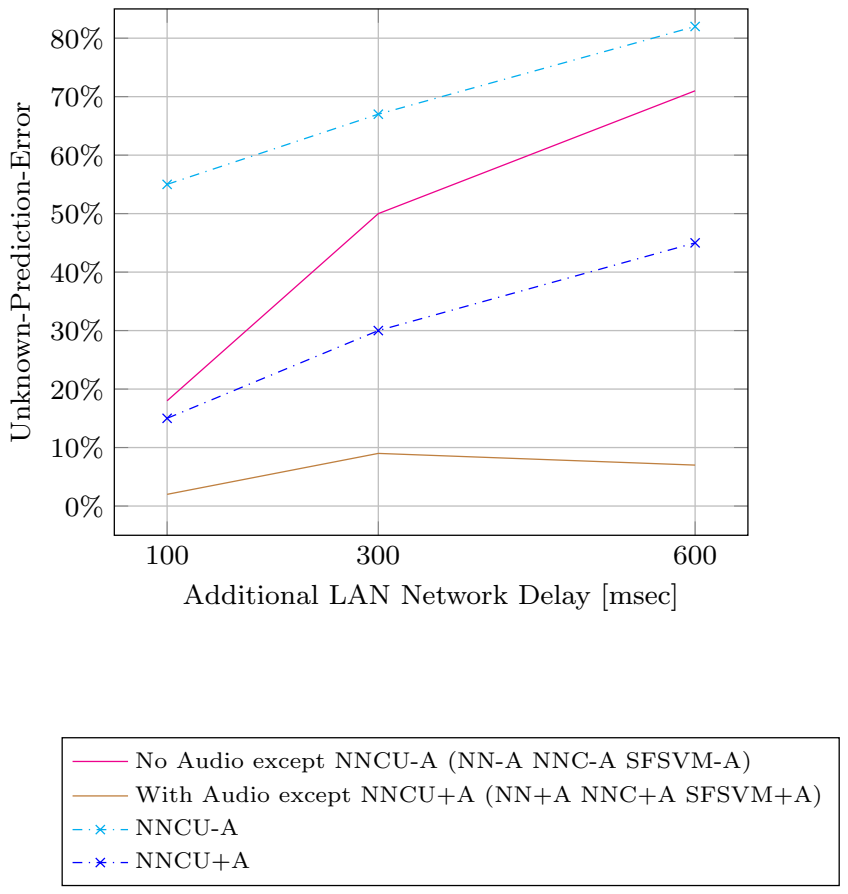

(j) Accuracy, 5 streams per video title in training (500 streams total)

(k) False-Prediction-Error, 5 streams per video title in training (500 streams total)

(l) Unknown-Prediction-Error, 5 streams per video title in training (500 streams total)

| | |
|---|---|
| NN: Nearest Neighbor | NNC: Nearest Neighbor to Class |
| NNCU: Nearest Neighbor to Class Unique | SFSVM: Similarities as Features Support Vector Machine |
| -A: without audio features | +A: with audio features |

Fig. 4: Accuracy, False-Prediction-Error and False-Unknown-Error results for different training data set sizes, different additional LAN network delay and different learning algorithms. Using the largest training dataset the SFSVM+A method outperformed all other methods and achieved accuracy of more than 80% even with severe network delays of 600msec. We conjecture that the learning phase and the usage of distance to class made this algorithm robust to changes in testing time. We can see that NNC+A accuracy was also high. Training with less streams (see next page), NNC+A and SFSVM+A were comparable and both outperformed all other algorithms. Like previous results the NNCU algorithm and not using audio features eliminated False-Prediction-Error. Out of all algorithms with low False-Prediciton-Error, NNCU+A accuracy was the best.

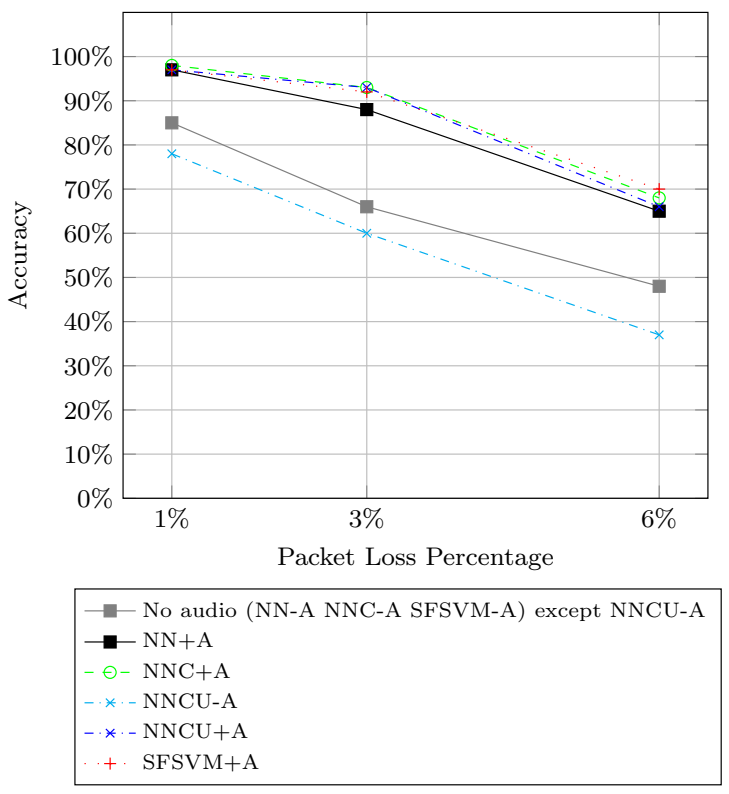

(a) Accuracy, 90 streams per video title in training (9000 streams total)

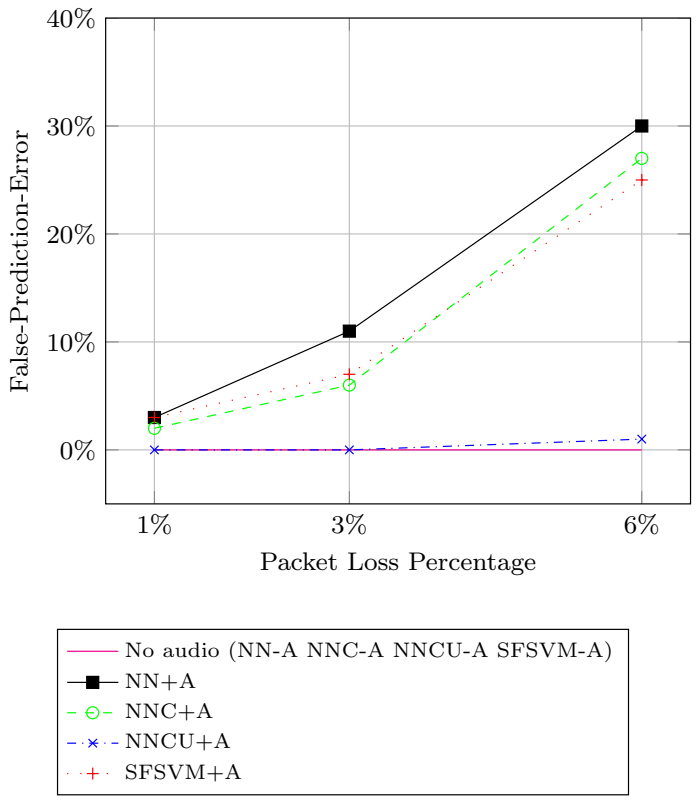

(b) False-Prediction-Error, 90 streams per video title in training (9000 streams total)

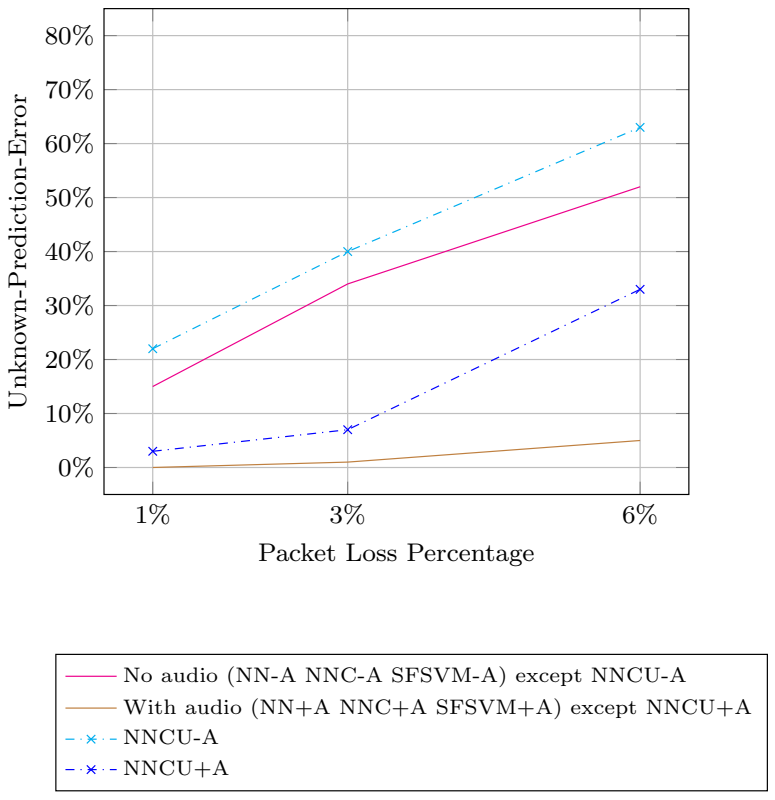

(c) Unknown-Prediction-Error, 90 streams per video title in training (9000 streams total)

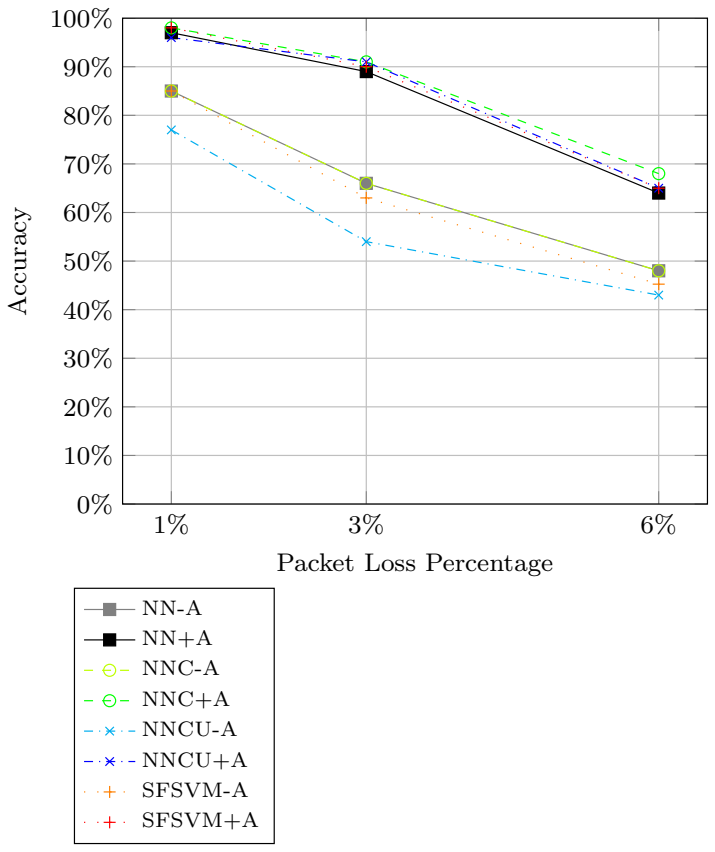

(d) Accuracy, 60 streams per video title in training (6000 streams total)

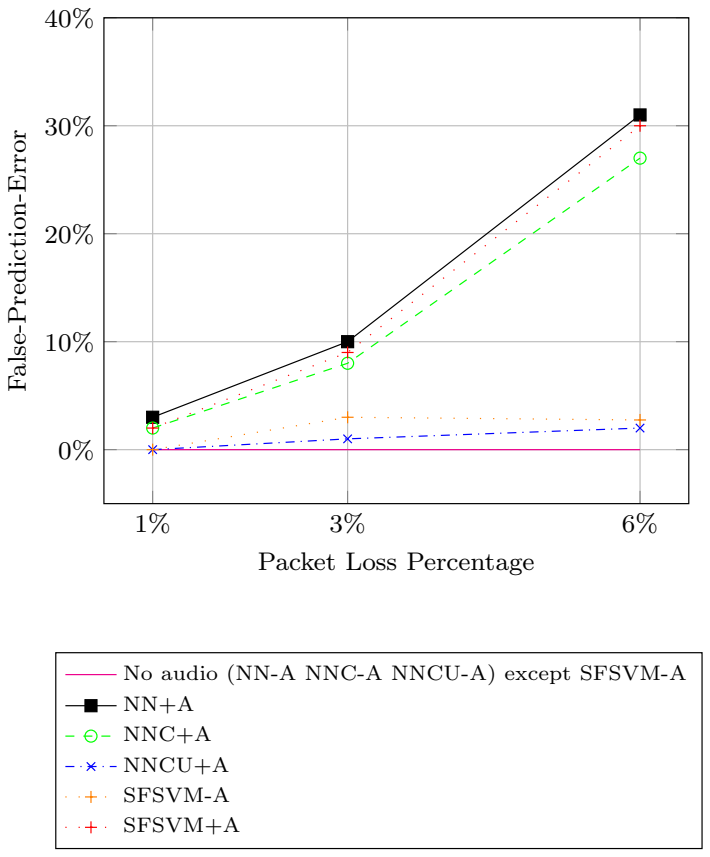

(e) False-Prediction-Error, 60 streams per video title in training (6000 streams total)

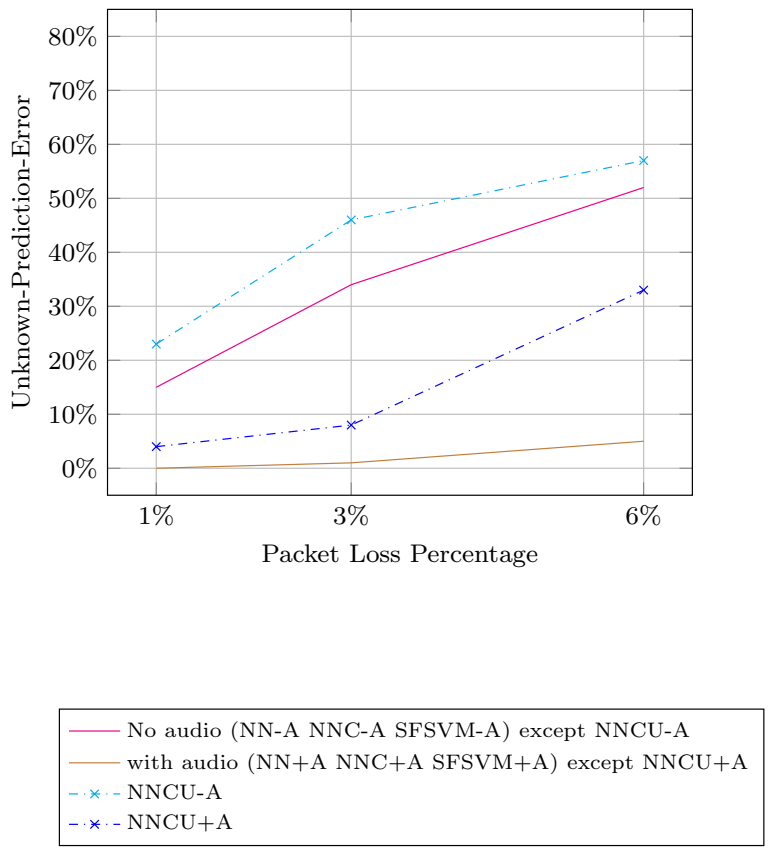

(f) Unknown-Prediction-Error, 60 streams per video title in training (6000 streams total)

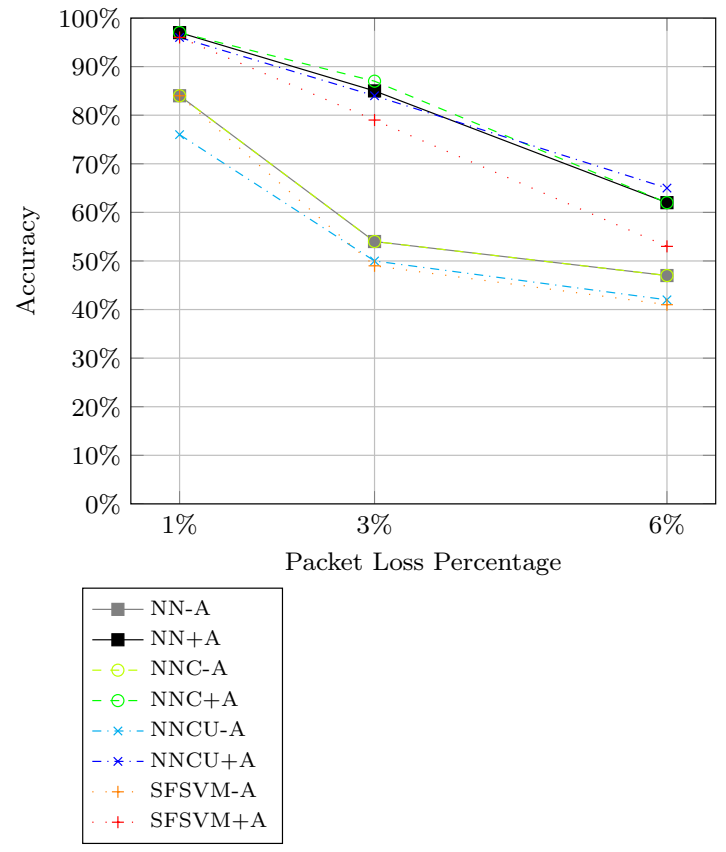

(g) Accuracy, 30 streams per video title in training (3000 streams total)

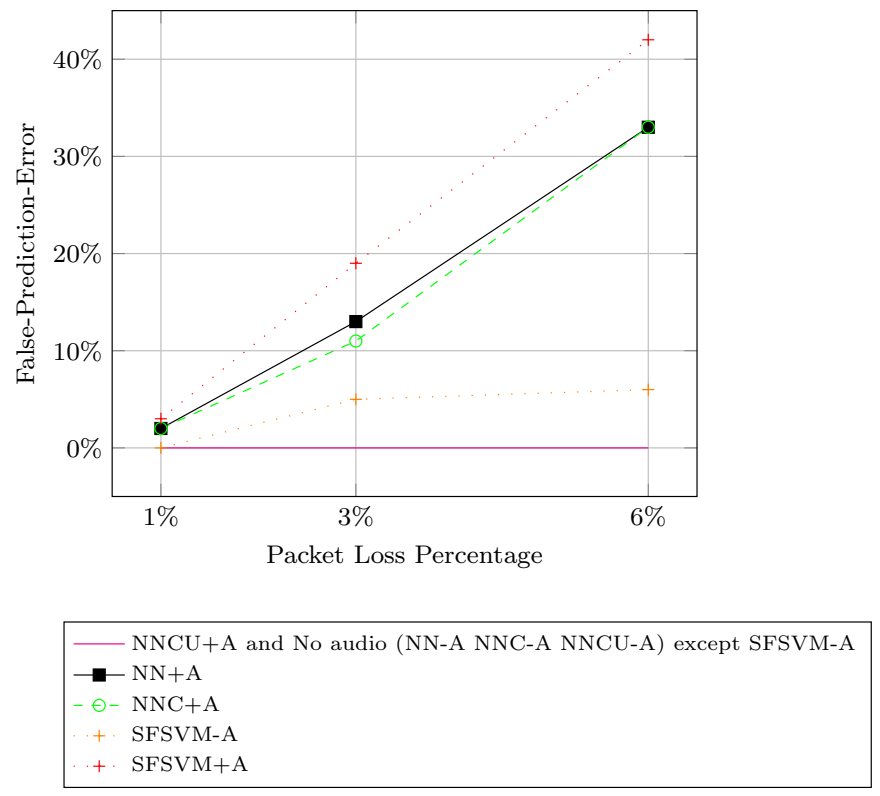

(h) False-Prediction-Error, 30 streams per video title in training (3000 streams total)

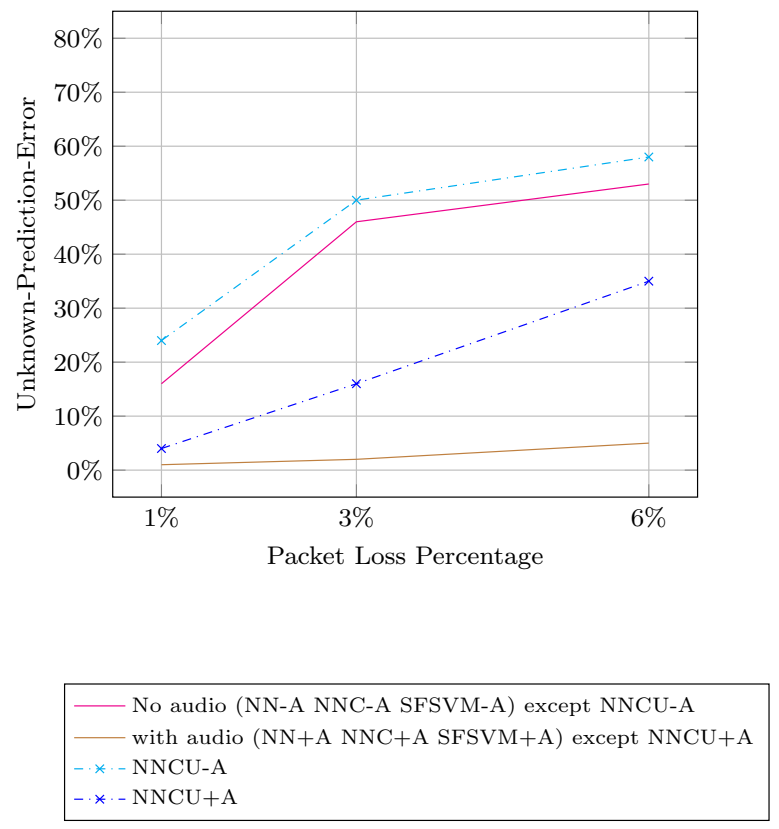

(i) Unknown-Prediction-Error, 30 streams per video title in training (3000 streams total)

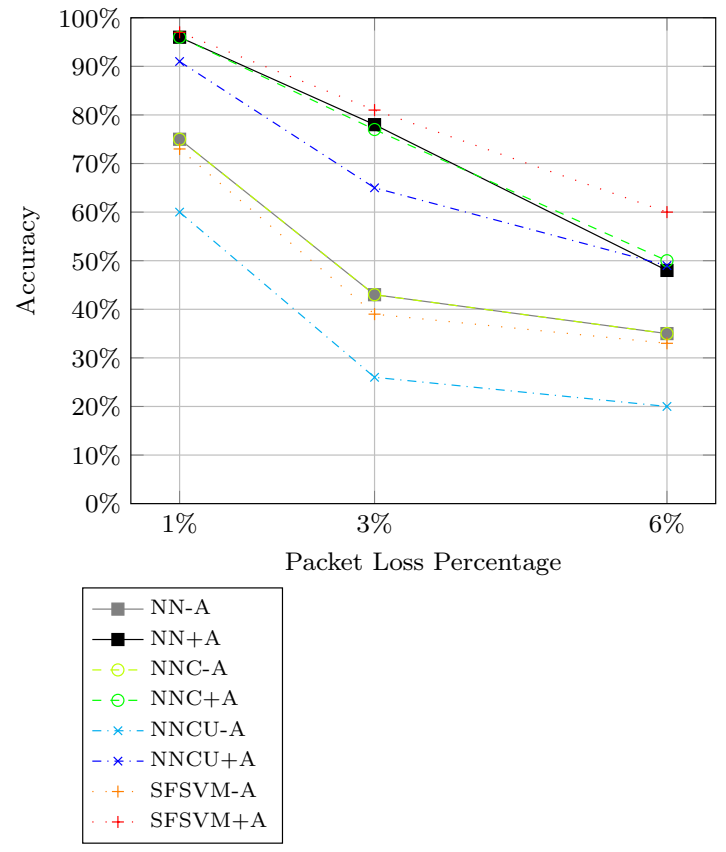

(j) Accuracy, 5 streams per video title in training (500 streams total)

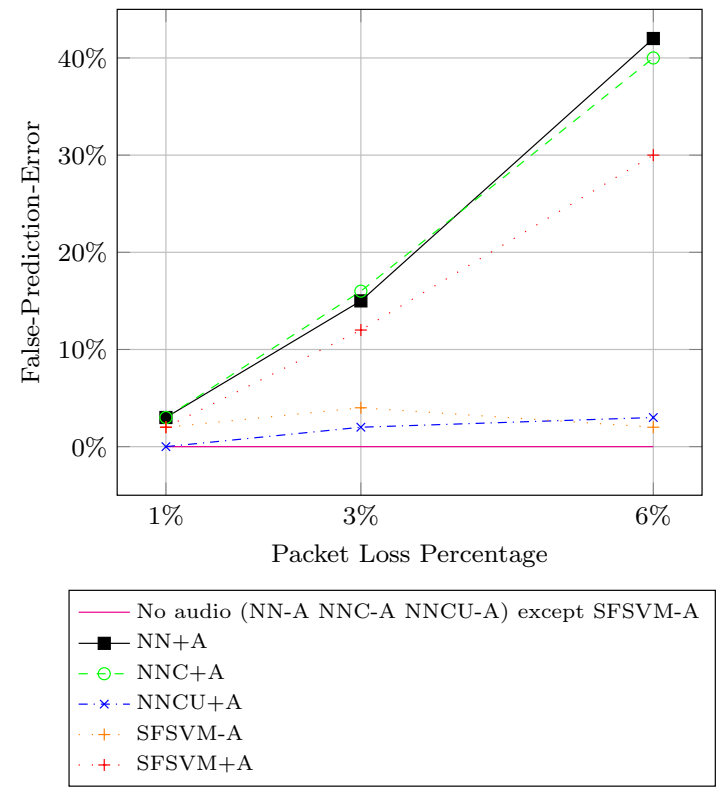

(k) False-Prediction-Error, 5 streams per video title in training (500 streams total)

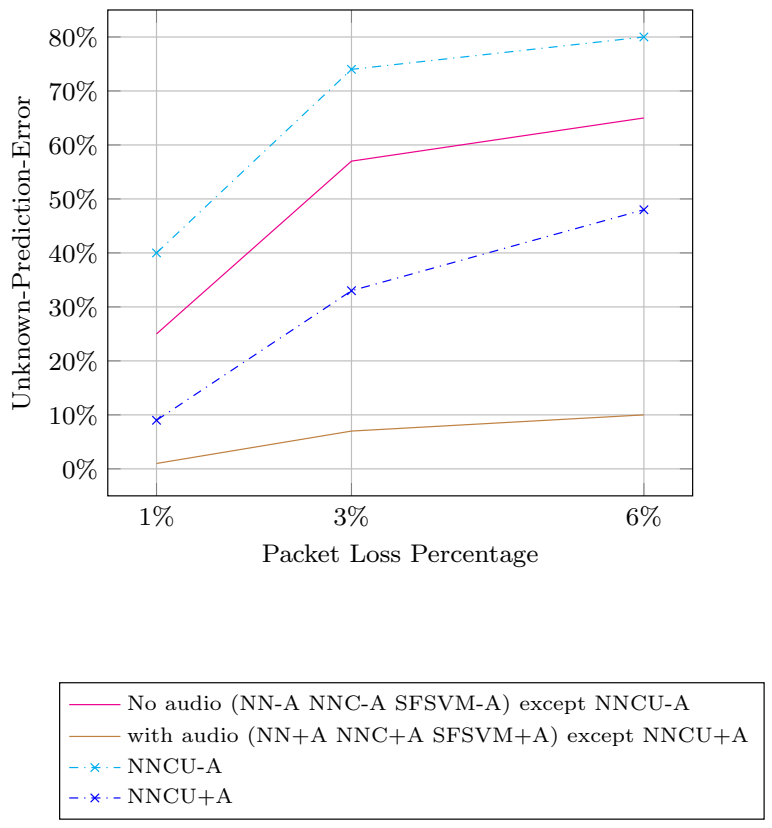

(l) Unknown-Prediction-Error, 5 streams per video title in training (500 streams total)

| | |
|---|---|
| NN: Nearest Neighbor | NNC: Nearest Neighbor to Class |
| NNCU: Nearest Neighbor to Class Unique | SFSVM: Similarities as Features Support Vector Machine |
| -A: without audio features | +A: with audio features |

Fig. 5: Accuracy, False-Prediction-Error and False-Unknown-Error results for different training data set sizes, different additional packet loss percentage and different learning algorithms. Using the largest training dataset the SFSVM+A method slightly outperformed all other methods and achieved accuracy of more than 70% even under severe packet loss of 6%. NNC+A, NNCU+A, NN+A accuracies were also good. Training with less streams resulted in generally reduced method accuracy. Similar to previous results the NNCU algorithm and not using audio features almost eliminated False-Prediction-Error. Again, out of all algorithms with low False-Prediciton-Error, NNCU+A accuracy was the best.